\documentclass[10pt,twocolumn,twoside]{IEEEtran}
\IEEEoverridecommandlockouts

\usepackage{cite}
\usepackage{tikz}
\usepackage{amsmath}
\usepackage{amsthm}
\newtheorem{remark}{Remark}
\usetikzlibrary{patterns,arrows.meta}
\usepackage{amssymb,amsfonts}
\usepackage{algorithmic}
\usepackage{graphicx}
\usepackage{textcomp}
\usepackage{balance}
\usepackage{xcolor}
\usepackage[utf8]{inputenc}
\usepackage{amssymb}

\usepackage{xcolor}

\newcommand{\review}[1]{#1}

\usepackage{amsthm}
\newcommand{\Ahmed}[1]{#1}
\def\BibTeX{{\rm B\kern-.05em{\sc i\kern-.025em b}\kern-.08em
    T\kern-.1667em\lower.7ex\hbox{E}\kern-.125emX}}
\usepackage{lipsum}

\title{\LARGE \bf 
Safe Reconnection Time for Large-Scale Data Center Loads: An Analytical Framework for Transient Stability Assessment 
}

\author{Ahmed Mesfer Alkhudaydi and Bai Cui
\vspace{-0.5cm}
 \thanks{ 
A. M. Alkhudaydi and B. Cui are with the Department of Electrical and Computer Engineering, Iowa State University, Ames, IA 50011 USA. Emails: {\tt \{ahmed92, baicui\}@iastate.edu}}
}

\begin{document}
\begingroup
\allowdisplaybreaks

\maketitle

\begin{abstract}
The rapid growth of large, power electronics-rich data center loads (DCLs) is creating new operational challenges for bulk power systems. A key risk arises when a DCL's uninterruptible power supply (UPS) disconnects the facility during voltage/frequency disturbances and then reconnects it while the bulk grid is still dynamically settling to a new equilibrium point. Poorly timed reconnection can amplify electromechanical oscillations, deepen frequency deviations, and lead to repeated connect-disconnect \emph{flapping}. \review{In this paper, we develop an analytical framework to characterize the \emph{safe reconnection windows} for large DCLs after a disturbance-induced disconnection that satisfy the prescribed limits. Using a model in the spirit of the classical single-machine infinite-bus system, we capture (i) swing dynamics during the disconnection interval and (ii) voltage-angle coupling at the load bus, which determines the electrical power step at reconnection under constant power load assumptions. Using the energy function method, we characterize the \emph{certified safe reconnection windows} such that the post-reconnection trajectory is guaranteed to remain within operational limits (rotor speed/angle/voltage).} Time-domain simulations validate the effectiveness of the proposed analytical approach. The results provide a simple physics-informed criterion that can be used to identify certified reconnection windows for large DCL facilities and inform UPS reconnection logic.

\end{abstract}



\section{Introduction}
Data center loads (DCLs) are becoming one of the fastest-growing classes of electricity demand, driven by cloud services and AI workloads. Recent international assessments highlight both the current scale and the potential for significant growth in DCL electricity consumption over this decade \cite{IEA_Electricity2024,IEA_EnergyAI}. This rapid expansion poses significant challenges for
transmission system operators (TSOs), particularly in island power systems and regions with high renewable
energy penetration which have limited inertia or constrained interconnection, where large step changes in demand can translate into material voltage and frequency excursions\cite{Saha2023RES_FrequencyBehaviour, ElengaBaningobera2025FFR_WAMS}.

Unlike many traditional industrial loads, DCLs exhibit unique operational characteristics that complicate power system
stability analysis. Modern DCLs are equipped with uninterruptible power supply (UPS) systems that provide fault
ride-through capability but also introduce complex disconnection and reconnection logic based on voltage and
frequency thresholds \cite{Guerrero2007,Guerrero2008,IEC62040_3_2021,IEEE1159_2019}. During grid disturbances, these protection systems can abruptly
disconnect hundreds of megawatts of load, causing substantial frequency excursions that trigger cascading
stability concerns~\cite{kundur_pssc}. From a transmission system perspective, the disturbance response is not limited to the initial disconnection. The reconnection moment can be equally (or more) consequential: if reconnection occurs while generator rotor dynamics are still settling, the resulting load step may (i) deepen the frequency nadir, (ii) increase rate of change of frequency (RoCoF), and/or (iii) excite electromechanical oscillations that push voltage or frequency back across the UPS trip thresholds, which is a phenomenon known as \emph{flapping} where the DCL repeatedly connects and
disconnects, further degrading system stability~\cite{jimenez_ruiz_milano_dc_transient_2025}.

Real-world incidents underscore the severity of this issue. In the all-island Irish transmission system, a fault in 2023 resulted in a 204~MW instantaneous drop in DCL demand, producing a RoCoF of 0.12~Hz/s and a frequency peak of 50.22~Hz~\cite{kerci_cigre_c4_10907_2024}. Ireland's position as an island system with limited interconnection capacity, together with rapidly growing DCL penetration, makes it
particularly vulnerable to such events. A 2021 CRU report cited a median-demand forecast in which data centers could account for 23\% of electricity demand in Ireland by 2030~\cite{cru_21_124}. 

Recent work in~\cite{jimenez_ruiz_milano_dc_transient_2025} developed a comprehensive DCL model incorporating UPS dynamics, cooling system induction motors, and time-varying computational loads, which provides valuable insights into DCLs behavior during faults. However, their work did not address the optimization of reconnection timing to prevent
post-fault instability. The determination of optimal reconnection time presents a challenging nonlinear optimization problem. The rotor angle and speed of synchronous generators evolve according to swing dynamics during the disconnection period, while the reconnection instant must satisfy both pre-connection constraints (adequate frequency and voltage recovery) and post-connection constraints (avoiding excessive oscillations that trigger
re-disconnection). Furthermore, the generator electrical output after DCL reconnection exhibits a strong nonlinear dependence on the generator rotor angle through voltage–angle coupling, which makes the solution more challenging~\cite{pai_energy_function}.

\review{To connect DCL dynamics with actionable operational decisions, we propose an analytical framework for characterizing certified safe reconnection windows following UPS-triggered disconnection, where the earliest certified reconnection time corresponds to the first point of the first certified window.} Our contributions are threefold: first, we develop a three-bus single-machine infinite-bus (SMIB) extension that retains voltage–angle coupling at the load bus while isolating the effect of reconnection timing; second, we develop a hybrid system representation of DCL disconnection/reconnection dynamics through explicit switching logic; third, we formulate a series of optimization problems for determining the certified safe reconnection windows, including the earliest certified reconnection time. The overall framework facilitates rapid computation of certified safe reconnection windows suitable for real-time operations.

The remainder of this paper is organized as follows. Section~\ref{sec:model} presents the system model, including the configuration of three-bus system, the swing equation, and the characteristics of the DCL. \review{Section~\ref{sec:problem} formalizes the problem of characterizing the certified safe reconnection windows of DCLs following a disturbance-induced disconnection. Section~\ref{sec:methodology} develops an energy function-based certification method to compute the certified safe reconnection windows, which includes the earliest certified reconnection time.} Section~\ref{sec:results} presents time-domain simulations of the proposed approach. Finally, Section~\ref{sec:Conclusion} concludes the paper.

\section{System Modeling}
\label{sec:model}
\subsection{A Simplified Three-Bus Power System Configuration}

We consider a simplified three-bus power system model, shown in Fig.~\ref{fig:threebus}, which extends the classical single-machine infinite-bus (SMIB) framework by explicitly introducing a dedicated DCL bus while preserving analytical tractability. Bus~1 represents a synchronous generator described by the second-order swing dynamics with an internal voltage $e\angle\delta$, where $\delta$ is the rotor angle referenced to the infinite bus. Bus~2 is the DCL bus, characterized by a constant power demand $(p_{\mathrm{DCL}}, q_{\mathrm{DCL}})$ and bus voltage $v\angle\phi$. Bus~3 is modeled as an infinite bus with fixed voltage $1\angle 0^\circ$, representing the external grid. The generator is connected to bus~2 through the transient reactance $x_d'$, which is suitable for stability at the first-swing and analysis of short-term disturbances~\cite{ieee_tf_standard_load_models_1995}. The transmission line connecting bus~2 and bus~3 has a reactance $x_l$ that varies according to the conditions of the system where $x_l^{\mathrm{pre}}$ in normal operation, $x_l^{\mathrm{f}}$ during fault, and $x_l^{\mathrm{post}}$ after a fault is cleared. 

Under steady-state conditions, bus 1, bus 2, and bus 3 are modeled as PV bus, PQ bus, and reference bus, respectively. For simplicity, we assume that the voltage magnitude at bus 1, the back emf of the synchronous generator, is constant, even during transients. \Ahmed{This assumption is standard in classical first-swing transient stability
analysis~\cite{kundur_pssc,sauer_pai_chow_2017}, where the generator
internal emf behind transient reactance is treated as approximately
constant during the electromechanical swing.  This approximation is most
appropriate when the study focuses on the first-swing response, the
generator remains within its reactive power capability, and the AVR or
excitation limiters do not dominate the voltage response.  Including
explicit AVR dynamics or fast reactive power support would change the
post-reconnection voltage trajectory at bus~2 and therefore affect the
DCL voltage bound constraints. A detailed treatment of these effects is beyond the scope of this work and is left for future study.} 

\begin{figure}[t!]
    \centering
\includegraphics[width=0.95\linewidth]{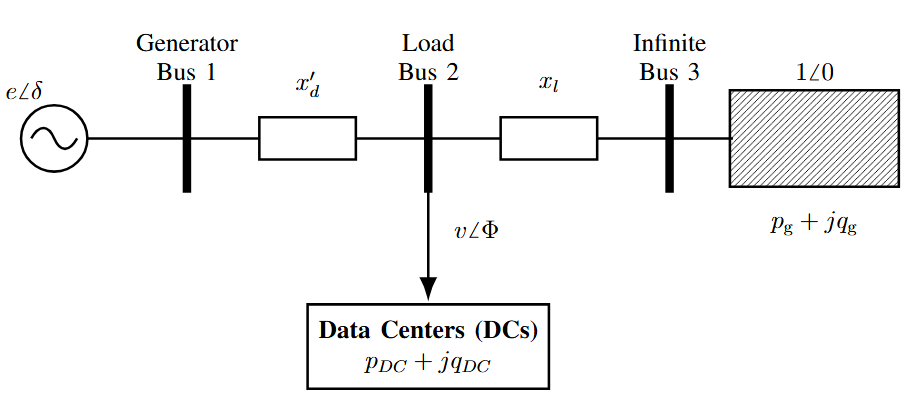}
    \caption{A simplified three-bus system}
    \label{fig:threebus}
\end{figure}

\subsection{Simulation Framework}
\label{subsec:sim_outline}

The dynamic behavior of the system is analyzed through four distinct stages: pre-fault steady state, fault-on dynamics, post-fault operation with the DCL disconnected, and reconnection. Each stage captures specific aspects of the interaction between the generator and the DCL during disturbances.

\review{We set fault clearing time as the origin, i.e., $t=0$, and denote
the fault duration by $t_f>0$. Therefore, the fault occurs at
$t=-t_f$. In other words, the system evolves under the fault-on dynamics duration $-t_f \le t < 0$,
and subsequently follows the DCL-disconnected waiting dynamics
before reconnection at $t=t_{\mathrm{re}}$.
All reconnection times are measured from fault clearing.}

\subsubsection{\textbf{Pre-fault State $(t<-t_f)$}}
The generator supplies the DCL and the infinite bus.  The system can be characterized by the steady state model as follows:
\begin{subequations}\label{eq:pf_ss}
\begin{align}
-p_{\mathrm{DCL}} &=
\frac{ev^{\mathrm{ss}}\sin(\phi^{\mathrm{ss}}-\delta^{\mathrm{ss}})}{x_d'}
+ \frac{v^{\mathrm{ss}}\sin\phi^{\mathrm{ss}}}{x_l^{\mathrm{pre}}}, \label{eq:pf_ss:a}\\
-q_{\mathrm{DCL}} &=
\frac{(v^{\mathrm{ss}})^2-ev^{\mathrm{ss}}\cos(\phi^{\mathrm{ss}}-\delta^{\mathrm{ss}})}{x_d'}
+ \frac{(v^{\mathrm{ss}})^2-v^{\mathrm{ss}}\cos\phi^{\mathrm{ss}}}{x_l^{\mathrm{pre}}},\label{eq:pf_ss:b}\\
p_g &=
\frac{ev^{\mathrm{ss}}\sin(\delta^{\mathrm{ss}}-\phi^{\mathrm{ss}})}{x_d'}. \label{eq:pf_ss:c}
\end{align}
\end{subequations}
Here, $p_{\mathrm{DCL}}, q_{\mathrm{DCL}}, p_g, x_d', x_l^{\mathrm{pre}},$ and $e$ are given parameters, while
$v^{\mathrm{ss}}, \phi^{\mathrm{ss}},$ and $\delta^{\mathrm{ss}}$ are unknown steady state variables.

Solving for $\sin (\delta^{\mathrm{ss}} - \phi^{\mathrm{ss}})$ in \eqref{eq:pf_ss:c} and $\sin \phi^{\mathrm{ss}}$ in \eqref{eq:pf_ss:a} leads to the following expressions:
\begin{subequations}\label{eq:pf_sines}
\begin{align}
\sin(\delta^{\mathrm{ss}}-\phi^{\mathrm{ss}}) &= \frac{p_g x_d'}{ev^{\mathrm{ss}}},\\
\sin\phi^{\mathrm{ss}} &= \frac{(p_g - p_{\mathrm{DCL}})\,x_l^{\mathrm{pre}}}{v^{\mathrm{ss}}}.
\end{align}
\end{subequations}

We can then replace $\cos (\delta^{\mathrm{ss}} - \phi^{\mathrm{ss}})$ and $\cos \phi^{\mathrm{ss}}$ in \eqref{eq:pf_ss:b} by solving for them in \eqref{eq:pf_sines}, which leads to:
\begin{multline}
-q_{\mathrm{DCL}} =
\left(\frac{1}{x_d'} + \frac{1}{x_l^{\mathrm{pre}}}\right)(v^{\mathrm{ss}})^2
- \frac{\sqrt{e^2(v^{\mathrm{ss}})^2 - p_g^2 x_d'^2}}{x_d'} \\
- \frac{\sqrt{(v^{\mathrm{ss}})^2 - (p_g - p_{\mathrm{DCL}})^2 (x_l^{\mathrm{pre}})^2}}{x_l^{\mathrm{pre}}},
\label{eq:quartic_v}
\end{multline}
where we assume $\delta^{\mathrm{ss}} - \phi, \phi \in [-\pi/2, \pi/2]$.

Equation \eqref{eq:quartic_v} is a quartic equation in $v^2$ that is cumbersome to solve directly. A simpler solution approach is to cast it in fixed-point form and apply fixed-point iteration:
\begin{multline}
v^{\mathrm{ss}}
= \frac{x_l^{\mathrm{pre}}}{x_d' + x_l^{\mathrm{pre}}}
\sqrt{e^2 - \frac{p_g^2 x_d'^2}{(v^{\mathrm{ss}})^2}}\\
 + \frac{x_d'}{x_d' + x_l^{\mathrm{pre}}}
\sqrt{1 - \frac{(p_g - p_{\mathrm{DCL}})^2 (x_l^{\mathrm{pre}})^2}{(v^{\mathrm{ss}})^2}} 
- \frac{x_d' x_l^{\mathrm{pre}}}{x_d' + x_l^{\mathrm{pre}}}\frac{q_{\mathrm{DCL}}}{v^{\mathrm{ss}}}.
\label{eq:fixed_point_form}
\end{multline}
After obtaining the steady-state voltage $v^{\mathrm{ss}}$, the phase angles $\delta^{\mathrm{ss}}$ and $\phi^{\mathrm{ss}}$ can be obtained from \eqref{eq:pf_sines}. 
\Ahmed{
\begin{remark}
The fixed-point iteration \eqref{eq:fixed_point_form} is well-defined
when the square root arguments remain nonnegative; convergence is
expected when the operating point is sufficiently far from voltage
collapse so that the corresponding map remains locally contractive.
\end{remark}
}
\subsubsection{\textbf{Fault On State $(-t_f\le t< 0)$}}
\review{At fault inception, $t = -t_f$, the modeled DCL grid demand is set to zero and the line reactance changes from $x_l^{\mathrm{pre}}$ to $x_l^f$. During $-t_f\le t< 0$, the swing dynamics are given by \eqref{eq:swing_pre_reconnect} with $x_l^{\mathrm{post}}$ replaced by $x_l^f$. The network reduces to an
equivalent two-bus system:}
\begin{subequations}\label{eq:fault_equiv}
\begin{align}
\tilde{e}_{\mathrm{eq}}^{\mathrm{f}}(\delta)
&= \frac{e\angle\delta / x_d'}{1/x_d' + 1/x_l^{\mathrm{f}}}
   + \frac{1/x_l^{\mathrm{f}}}{1/x_d' + 1/x_l^{\mathrm{f}}} \notag\\
&= \frac{(x_l^{\mathrm{f}} e\cos\delta + x_d') + j x_l^{\mathrm{f}} e\sin\delta}{x_d' + x_l^{\mathrm{f}}},\\
x_{\mathrm{eq}}^{\mathrm{f}}
&= \frac{1}{1/x_d' + 1/x_l^{\mathrm{f}}}
 = \frac{x_d' x_l^{\mathrm{f}}}{x_d' + x_l^{\mathrm{f}}}.
\end{align}
\end{subequations}

\review{With zero DCL grid demand, the voltage at bus 2 is the same as $\tilde{e}_{\mathrm{eq}}^{\mathrm{f}}(\delta)$:}
\begin{equation}
v^{\mathrm{f}}(\delta)
= \left|\tilde{e}_{\mathrm{eq}}^{\mathrm{f}}(\delta)\right|.
\label{eq:v_fault}
\end{equation}
\begin{equation}
\phi^{\mathrm{f}}(\delta)
=
\angle \tilde{e}_{\mathrm{eq}}^{\mathrm{f}}(\delta).
\label{eq:phi_fault}
\end{equation}


\subsubsection{\textbf{Fault Clearing to Reconnection State $(0\le t<t_{\mathrm{re}})$}}

\review{At $t=0$, the fault is cleared and the line reactance changes
to $x_l^{\mathrm{post}}$. The generator angle and speed remain
continuous across clearing, so the waiting trajectory starts
from the terminal fault-on state:
\begin{equation}
\delta(0^+)=\delta(0^-), \qquad
\omega(0^+)=\omega(0^-).
\end{equation}
During $0 \le t < t_{\mathrm{re}}$, the DCL remains disconnected
and the generator evolves according to the waiting dynamics
in~\eqref{eq:swing_pre_reconnect}, toward the no-load equilibrium defined in~\eqref{eq:no_load_equilibrium} below}:
\begin{subequations}\label{eq:no_load_equilibrium}
\begin{align}
p_g &= \frac{e \sin (\delta^{\mathrm{no\mbox{-}load}}_{\mathrm{eq}})}{x'_d + x_l^{\mathrm{post}}}\,,
\label{eq:no_load_eq}\\
\delta^{\mathrm{no\mbox{-}load}}_{\mathrm{eq}}
&= \arcsin\!\left(\frac{p_g\,(x'_d + x_l^{\mathrm{post}})}{e}\right).
\label{eq:delta_no_load}
\end{align}
\end{subequations}

The system dynamics are therefore governed by the swing equations
\begin{subequations}\label{eq:swing_pre_reconnect}
\begin{align}
\dot{\delta} &= \omega,\\
M\dot{\omega} &= p_g - \frac{e}{x_d' + x_l^{\mathrm{post}}}\sin\delta - D\omega,
\end{align}
\end{subequations}
where $M$ is the generator inertia constant and $D$ is the damping coefficient.

\subsubsection{\textbf{Reconnection State $(t\ge t_{\mathrm{re}})$}}
\review{After reconnection at a candidate reconnection time $t_{\mathrm{re}}$, the post-fault equivalent system parameters are}
\begin{subequations}\label{eq:post_equiv}
\begin{align}
\tilde{e}_{\mathrm{eq}}^{\mathrm{post}}(\delta)
&= \frac{e\angle\delta / x_d'}{1/x_d' + 1/x_l^{\mathrm{post}}}
   + \frac{1/x_l^{\mathrm{post}}}{1/x_d' + 1/x_l^{\mathrm{post}}} \notag\\
&= \frac{(x_l^{\mathrm{post}} e\cos\delta + x_d') + j x_l^{\mathrm{post}} e\sin\delta}{x_d' + x_l^{\mathrm{post}}},\\
x_{\mathrm{eq}}^{\mathrm{post}}
&= \frac{1}{1/x_d' + 1/x_l^{\mathrm{post}}}
 = \frac{x_d' x_l^{\mathrm{post}}}{x_d' + x_l^{\mathrm{post}}}.
\end{align}
\end{subequations}

The bus~2 voltage magnitude and phase angle are
\begin{subequations}\label{eq:bus2_post}
\begin{align}
 v^{\mathrm{post}}(\delta)
&= \sqrt{
\frac{
|\tilde{e}_{\mathrm{eq}}^{\mathrm{post}}(\delta)|^{2}
\review{-} 2x_{\mathrm{eq}}^{\mathrm{post}}q_{\mathrm{DCL}}
+ \sqrt{\Delta^{\mathrm{post}}(\delta)}
}{2}}
,\label{eq:bus2_post:a}\\
\phi^{\mathrm{post}}(\delta)
&= \angle \tilde{e}_{\mathrm{eq}}^{\mathrm{post}}(\delta) + \angle\!\bigl(\left[v^{\mathrm{post}}(\delta)\right]^2 + \notag \\
& \hspace{1in}  x_{\mathrm{eq}}^{\mathrm{post}}q_{\mathrm{DCL}}
- jx_{\mathrm{eq}}^{\mathrm{post}}p_{\mathrm{DCL}}\bigr), \label{eq:bus2_post:b}
\end{align}
\end{subequations}
where the discriminant is given via \begin{multline*} \Delta^{\mathrm{post}}(\delta) = \left(|\tilde{e}_{\mathrm{eq}}^{\mathrm{post}}(\delta)|^{2} \review{-}  2x_{\mathrm{eq}}^{\mathrm{post}}q_{\mathrm{DCL}}\right)^{2} - \\
4\left(x_{\mathrm{eq}}^{\mathrm{post}}\right)^{2} \left(p_{\mathrm{DCL}}^{2}+q_{\mathrm{DCL}}^{2}\right). \label{eq:disc_reconnect} \end{multline*}
The post-reconnection system dynamics are governed by:
\begin{subequations}\label{eq:swing_post}
\begin{align}
\dot{\delta} &= \omega,\\
M\dot{\omega} &= p_g - \frac{e v^{\mathrm{post}}(\delta)}{x_d'}
\sin\!\big(\delta - \phi^{\mathrm{post}}(\delta)\big) - D\omega.
\end{align}
\end{subequations}

The reconnection dynamic model above captures the transient following DCL reconnection. The interdependence between $\delta$, $v^{\mathrm{post}}(\delta)$, and $\phi^{\mathrm{post}}(\delta)$ in \eqref{eq:bus2_post:a}-\eqref{eq:bus2_post:b} indicates the coupling between generator dynamics and load voltage recovery during the reconnection process. \review{The objective of the subsequent methodology is to determine the set of reconnection times for which the post-reconnection trajectory satisfies the prescribed transient security constraints. As shown later, this set may consist of one or more disjoint certified safe reconnection windows.}

\section{Problem Formulation}
\label{sec:problem}

\review{In this section, we develop a direct method to characterize the certified safe reconnection windows for the DCL. These windows identify the reconnection times for which the post-reconnection trajectory is guaranteed to remain stable and satisfy the prescribed operational protection limits. The earliest certified reconnection time $t_{\mathrm{re}}^\star$ is defined as the first point of the first certified reconnection window.}

Let the generator swing state be
$x(t):=[\delta(t), \omega(t)]^\top$ and denote the post-reconnection (DCL-connected) stable equilibrium by
$(\delta_{\mathrm{eq}}^{\mathrm{load}},0)$. \review{Here, the reconnection time $t_{\mathrm{re}}$ acts as the switching time that determines the state at reconnection $x(t_{\mathrm{re}})$, which serves as the initial condition for the post-reconnection dynamics. The objective is to identify the subset of reconnection times for which this initial condition is certified to be safe.}

\subsection{Hybrid Evolution and Waiting Trajectory}
\label{subsec:hybrid}
The reconnection time \review{$t_{\mathrm{re}} > 0$} determines the switching instant 
between pre- and post-reconnection dynamics.

During the waiting interval $0\le t<t_{\mathrm{re}}$, the state $x(t)$ evolves under the pre-reconnection swing dynamics
(DCL-disconnected), cf.\ \eqref{eq:swing_pre_reconnect}.
\review{At $t=t_{\mathrm{re}}$,} the DCL is reconnected and the post-reconnection dynamics apply, cf.\ \eqref{eq:swing_post}.
This mirrors the role of fault clearing time in classical transient stability analysis where changing $t_{\mathrm{re}}$ changes the
switching state $x(t_{\mathrm{re}})$ and thus the subsequent trajectory.

\subsection{Certified Safe Initial Condition Set for Reconnection}
\label{subsec:safeset}
\review{ We adopt a cascaded formulation in which a set of post-reconnection initial conditions that are guaranteed to be safe is first characterized. The certified reconnection windows are then obtained as the time intervals during which the waiting trajectory evolves inside this set. The earliest certified reconnection time
$t_{\mathrm{re}}^\star$ is the first point of the first certified reconnection window.}

\subsubsection{Protection band in the $(\delta,\omega)$ plane}

Operational protection limits on rotor angle and frequency deviations are defined with respect to the post-reconnection equilibrium $(\delta_{\mathrm{eq}}^{\mathrm{load}},0)$ as
\begin{equation}
|\delta - \delta_{\mathrm{eq}}^{\mathrm{load}}| \le \delta_{\max}, \quad
|\Delta\omega| \le \omega_{\max}.
\label{eq:angle_freq_bounds}
\end{equation}


\review{These bounds are prescribed rotor-angle and rotor-speed limits for the reduced model, where $\Delta\omega = \omega = \dot{\delta}$.}

\subsubsection{Voltage bounds}
In the reduced order model, the voltage magnitude at bus 2 after reconnection is determined algebraically from the
post-reconnection network reduction and the constant power DCL model, which is given as $v^{\mathrm{post}}(\delta)$ in \eqref{eq:bus2_post}.
This voltage must satisfy operational bounds as
\begin{equation}
v_{\min} \le v^{\mathrm{post}}(\delta) \le v_{\max}.
\label{eq:v_post_bounds}
\end{equation}


Although $v^{\mathrm{post}}(\delta)$ depends only on $\delta$ as shown \eqref{eq:bus2_post}, the voltage constraint must be explicitly enforced, as the nonlinear voltage-angle coupling does not guarantee the satisfaction of \eqref{eq:v_post_bounds} solely from the rotor angle limits. Therefore, we define the admissible reconnection set as the set of post-reconnection initial conditions that
satisfy both the protection band and the voltage admissibility conditions as
\begin{equation}
\mathcal{S}_{\mathrm{adm}} := \left\{(\delta,\Delta\omega) \in \mathbb{R}^2 \;\Big|\;
\begin{aligned}
&|\delta - \delta_{\mathrm{eq}}^{\mathrm{load}}| \leq \delta_{\max}, \\
&|\Delta\omega| \leq \omega_{\max}, \\
&v_{\min} \leq v^{\mathrm{post}}(\delta) \leq v_{\max}
\end{aligned}
\right\}.
\label{eq:S_adm}
\end{equation}

The proposed method excludes instantaneous violations of the modeled limits. In practice, however, protection relays typically require the excursion to persist for a specific duration (on the order of cycles) before triggering a trip. As a result, the certified safe reconnection time can be conservative relative to delayed tripping for the same monitored quantities and thresholds, and incorporating a time delay coordination principle into the threshold analysis serves as a natural extension of this framework.

\subsection{Certified Safe Reconnection Windows}
\label{subsec:first_entry}
Let $x^{-}(t)$ denote the waiting trajectory generated by \eqref{eq:swing_pre_reconnect} with the DCL disconnected.
\review{Due to the oscillatory swing dynamics, the waiting trajectory $x^{-}(t)$ may enter and exit the certified safe set multiple times. Consequently, certified reconnection is not generally characterized by a single threshold time after which all later reconnection instants remain safe. Instead, the waiting trajectory may intersect the certified safe set over one or more disjoint time intervals, referred to as certified safe reconnection windows.}

\review{To formalize this characterization, we introduce the concept of a certified safe reconnection set $\mathcal{S}_{\mathrm{safe}}\subseteq\mathcal{S}_{\mathrm{adm}}$, which will be constructed in Section~\ref{sec:methodology} using energy function analysis. The certified reconnection windows are then obtained as the time intervals during which the waiting trajectory evolves inside $\mathcal{S}_{\mathrm{safe}}$.}

\Ahmed{In this hybrid framework, the system can be represented as a hybrid automaton with two 
discrete modes. \emph{Mode 1} (DCL-disconnected): the state $x(t)$ evolves 
under \eqref{eq:swing_pre_reconnect}. \emph{Mode 2} 
(DCL-connected): the state evolves under \eqref{eq:swing_post} with 
guard condition $x^{-}(t_{\mathrm{re}}) \in \mathcal{S}_{\mathrm{safe}}$. 
The reset map at the switching instant $t = t_{\mathrm{re}}$ is 
the identity, since reconnection does not introduce a jump in $(\delta, \omega)$.} Fig.~\ref{fig:phaseplane} illustrates the phase plane $(\delta,\omega)$. During the waiting interval, the state evolves
according to~\eqref{eq:swing_pre_reconnect}. Reconnection is permitted only when the trajectory reaches a switching state
that is certified safe for post-reconnection dynamics and sufficiently settled under the dissipating pre-reconnection
energy $E_{\mathrm{pre}}(t)$.

\begin{figure}[t!]
\centering
\includegraphics[width=\linewidth]{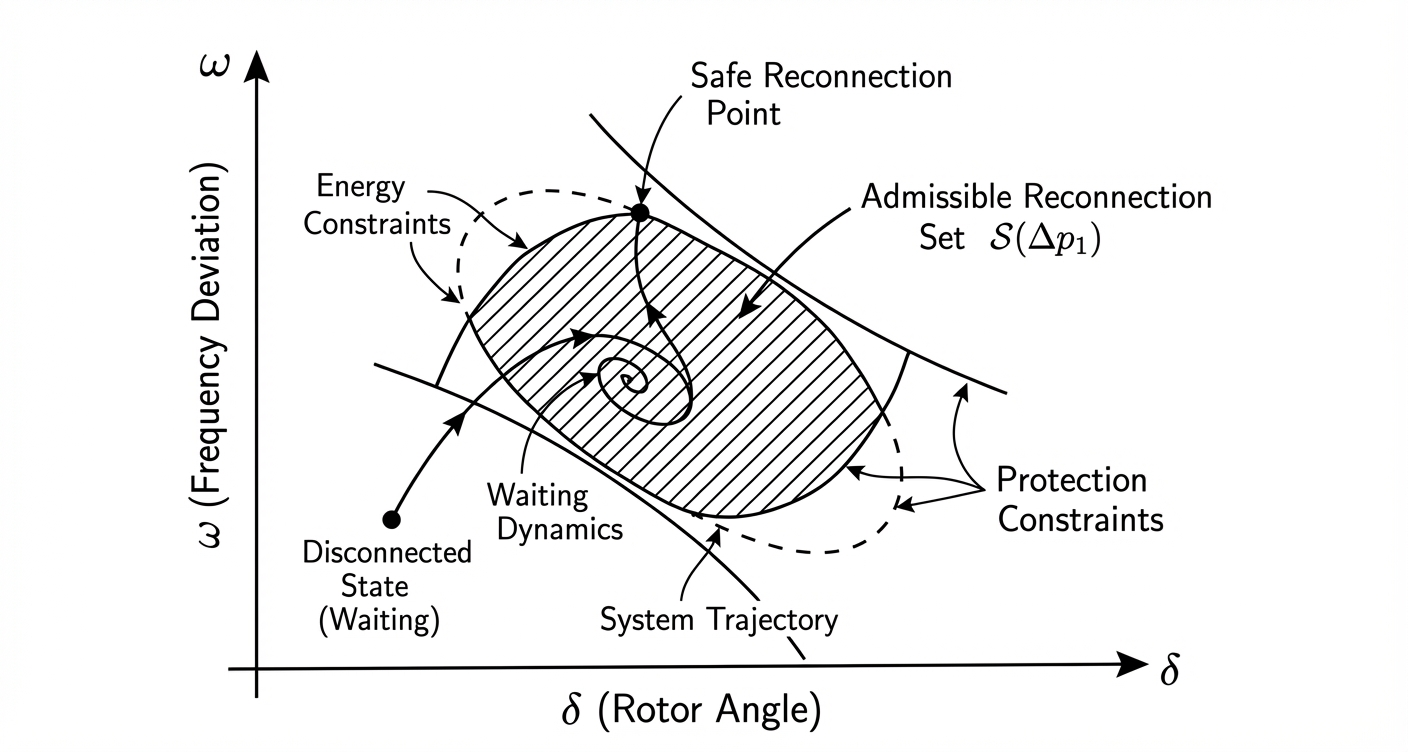}
\caption{Geometric interpretation in the $(\delta,\omega)$ plane: the waiting trajectory $x^{-}(t)$ evolves under
\eqref{eq:swing_pre_reconnect} \review{until it enters and evolves within the certified safe reconnection set $\mathcal{S}_{\mathrm{safe}}$, which produces one or more certified reconnection windows.}}
\label{fig:phaseplane}
\end{figure}

\section{Safe Reconnection Certification via Energy Function}
\label{sec:methodology}

Based on the dynamic model established in Section~\ref{sec:model}, this section develops an energy-based certification framework for the safe reconnection of a DCL. A reconnection is certified as safe if its post-reconnection trajectory remains within the specified admissible region. To address the topological change induced by reconnection, we employ dual-energy functions for the pre- and post-reconnection phases \cite{sauer_pai_chow_2017}.

\subsection{Pre-Reconnection Energy Function}
\label{subsec:pre_energy}

During the waiting interval $0 < t < t_{\mathrm{re}}$, the DCL is disconnected and the generator dynamics follows the no-load swing equation \eqref{eq:swing_pre_reconnect}. Let $\delta^{\mathrm{no\text{-}load}}_{\mathrm{eq}}$ denote the corresponding stable equilibrium satisfying $p_e^{\mathrm{no\text{-}load}}(\delta_{\mathrm{eq}}^{\mathrm{no\text{-}load}})=p_g$. The pre-reconnection energy function is defined as
\begin{equation}
E_{\mathrm{pre}}(\delta,\omega)
=
\frac{1}{2}M \omega^2
+
\underbrace{\int_{\delta^{\mathrm{no\text{-}load}}_{\mathrm{eq}}}^{\delta}
\Bigl[
\frac{e}{x'_d+x_l^{\mathrm{post}}}\sin\theta - p_g
\Bigr]\,d\theta}_{\Phi_{\mathrm{pre}}(\delta)}.
\label{eq:pre_energy}
\end{equation}

Along the trajectories of the no-load dynamics, the time derivative satisfies $\dot E_{\mathrm{pre}} = -D\omega^2 \le 0 $, which means that all sublevel sets $\Omega_{\mathrm{pre}}(c) := \{(\delta,\omega) : E_{\mathrm{pre}} \le c\}$ are a forward invariant set. 

\begin{proof}
By the chain rule and system dynamics \eqref{eq:swing_pre_reconnect},
\begin{align*}
\frac{d}{dt}E_{\mathrm{pre}}
&= M\omega\dot{\omega} + \frac{\partial\Phi_{\mathrm{pre}}}{\partial\delta}\dot{\delta}
= M\omega\dot{\omega} + [p_e^{\mathrm{no\text{-}load}}(\delta) - p_g]\omega \notag\\
&= \omega[p_g - p_e^{\mathrm{no\text{-}load}}(\delta) - D\omega] + [p_e^{\mathrm{no\text{-}load}}(\delta) - p_g]\omega\\
&=-D\omega^2.
\end{align*}
Since $D \geq 0$, we have $\dot{E}_{\mathrm{pre}} \leq 0$. Forward invariance of sublevel sets $\{E_{\mathrm{pre}} \leq c\}$ follows immediately from the non-increasing property.
\end{proof}

Let the pre-reconnection protection admissible region be
\begin{multline}
\mathcal B
:=
\Bigl\{
(\delta,\omega)\in\mathbb R^2:
\delta_{\mathrm{eq}}^{\mathrm{load}} - \delta_{\max} \le \delta \le \delta_{\mathrm{eq}}^{\mathrm{load}} + \delta_{\max}, \\
 -\omega_{\max} \le \omega\le\omega_{\max}
\Bigr\}.
\label{eq:D_box}
\end{multline}
Denote the pre-reconnection vector field via
\begin{equation}
f_{\mathrm{pre}}(\delta,\omega)
=
\begin{bmatrix}
\omega \\
\frac{1}{M}\bigl(p_g-p_e^{\mathrm{no\mbox{-}load}}(\delta)-D\omega\bigr)
\end{bmatrix}.
\label{eq:fpre}
\end{equation}
We define the outward-pointing portion of the boundary as
\begin{equation}
\partial\mathcal{B}_{\mathrm{out}}
:=
\bigl\{x\in \partial\mathcal{B}:\ n(x)^\top f_{\mathrm{pre}}(x) \ge 0\bigr\},
\label{eq:outward_boundary_def}
\end{equation}
\review{where $n(x)$ is an outward face normal; at a corner, the test is applied to each adjacent face and the point is included if any test holds.}

Because $E_{\mathrm{pre}}$ is non-increasing, a trajectory starting from $x_0:=(\delta_0,\omega_0)$
can only reach states whose energy is less than or equal to $E_{\mathrm{pre}}(x_0)$.
Therefore, a sufficient no-escape condition is that the initial energy lies strictly below
the minimum energy of all outward-pointing exit points as
\begin{equation}
E_{\mathrm{pre}}(x_0)
<
\inf_{x\in\partial\mathcal{B}_{\mathrm{out}}} E_{\mathrm{pre}}(x).
\label{eq:no_escape_condition}
\end{equation}
This motivates the following robust  critical-energy optimization as 
\begin{equation}
E_{\mathrm{crit,pre}}
:=
\inf_{x\in\partial\mathcal{B}_{\mathrm{out}}} E_{\mathrm{pre}}(x).
\label{eq:P1_robust}
\end{equation}

Let $\partial\mathcal{B}_{\mathrm{out}}$ be decomposed into four facewise 
subsets corresponding to the faces of $\mathcal{B}$. For each face, the 
energy barrier is the infimum of $E_{\mathrm{pre}}$ over the outward-pointing or tangential
region, which is defined as follows
\begin{subequations}
\label{eq:face_barriers}
\begin{align}
E_{\delta^+}
&:=
\inf_{(\delta_{\mathrm{eq}}^{\mathrm{load}} + \delta_{\max},\omega)\in\partial\mathcal{B}_{\mathrm{out}}} \; \bigl\{E_{\mathrm{pre}}(\delta_{\mathrm{eq}}^{\mathrm{load}} + \delta_{\max},\omega)\;: \notag \\
& \hspace{2.3in} \omega \ge 0\bigr\},
\label{eq:Edeltaplus}\\
E_{\delta^-}
&:=
\inf_{(\delta_{\mathrm{eq}}^{\mathrm{load}} - \delta_{\max},\omega)\in\partial\mathcal{B}_{\mathrm{out}}}\; \bigl\{E_{\mathrm{pre}}(\delta_{\mathrm{eq}}^{\mathrm{load}} - \delta_{\max},\omega): \notag \\
& \hspace{2.3in} \omega \le 0\bigr\},
\label{eq:Edeltaminus}\\
E_{\omega^+}
&:=
\inf_{(\delta,\omega_{\max})\in\partial\mathcal{B}_{\mathrm{out}}}\;
\bigl\{E_{\mathrm{pre}}(\delta,\omega_{\max})\;: \notag \\ 
& \hspace{1.8in}\dot{\omega}(\delta,\omega_{\max}) \ge 0\bigr\},
\label{eq:Eomegaplus}\\
E_{\omega^-}
&:=
\inf_{(\delta,-\omega_{\max})\in\partial\mathcal{B}_{\mathrm{out}}}\;
\bigl\{E_{\mathrm{pre}}(\delta,-\omega_{\max})\;: \notag\\
& \hspace{1.68in} \dot{\omega}(\delta, -\omega_{\max}) \le 0\bigr\}.
\label{eq:Eomegaminus}
\end{align}
\end{subequations}

The minimum barrier over all four faces defines the pre-reconnection critical energy as 
\begin{equation}
E_{\mathrm{crit,pre}} = \min\left\{E_{\delta^+},\,E_{\delta^-},\,E_{\omega^+},\,E_{\omega^-}\right\}.
\label{eq:critical_energy}
\end{equation}

Therefore, the certified pre-reconnection safe set is the energy sublevel set
\begin{equation}
\mathcal{S}_{\mathrm{pre}} := \left\{(\delta,\omega)\in\mathcal{B} : 
E_{\mathrm{pre}}(\delta,\omega) < E_{\mathrm{crit,pre}}\right\}.
\label{eq:pre_safe_set}
\end{equation}
By construction, any trajectory originating in $\mathcal{S}_{\mathrm{pre}}$ 
satisfies $E_{\mathrm{pre}}(x(t)) \leq E_{\mathrm{pre}}(x_0) < E_{\mathrm{crit,pre}}$ 
for all $t \geq 0$. Consequently, it cannot reach $\partial\mathcal{B}_{\mathrm{out}}$, 
as every point on $\partial\mathcal{B}_{\mathrm{out}}$ has energy at least 
$E_{\mathrm{crit,pre}}$.

\subsection{Post-Reconnection Energy Function}
\label{subsec:post_energy}

At $t=t_{\mathrm{re}}$, the DCL is reconnected and the dynamics switches to the load-connected system with a new equilibrium  $\delta^{\text{load}}_{\text{eq}}$. The post-reconnection energy function $E_{\mathrm{post}}(\delta, \omega)$ characterizes the ability of the system to absorb the transient induced by the load step as
\begin{multline}
E_{\mathrm{post}}(\delta,\omega)
=
\frac{1}{2}M \omega^2\\ 
+
\int_{\delta^{\mathrm{load}}_{\mathrm{eq}}}^{\delta}
\underbrace{\Bigl[
\frac{e\,v^{\mathrm{post}}(\theta)}{x'_d}
\sin\!\bigl(\theta-\varphi^{\mathrm{post}}(\theta)\bigr)}_{p_e^{\mathrm{load}}(\theta)}-p_g
\Bigr]\,d\theta.
\label{eq:post_energy}
\end{multline}
Along the trajectories of the post-reconnection dynamics, the time derivative satisfies $\dot{E}_{\mathrm{post}} = -D\omega^2 \leq 0$, \review{so energy is nonincreasing while the solution remains on the adopted voltage branch.}

To determine the post-reconnection critical energy $E_{\mathrm{crit,post}}$, a boundary sampling approach is employed. The critical energy is defined as the infimum of post-reconnection energy over the admissible boundary points associated with escape from the admissible operating region
\begin{equation}
E_{\mathrm{crit,post}}
:=
\inf_{x \in \partial\mathcal{S}_{\mathrm{adm}}^{\mathrm{out}}} E_{\mathrm{post}}(x),
\label{eq:Ecrit_post}
\end{equation}
where $\partial\mathcal{S}_{\mathrm{adm}}^{\mathrm{out}}$ denotes the outward-pointing or tangential portion of the admissible boundary. This boundary includes points where either the rotor angle deviation reaches its limit, the frequency deviation reaches its limit, or the post-reconnection voltage constraint becomes active.

A numerical estimate of $E_{\mathrm{crit,post}}$ is computed by dense sampling of the admissible boundary and evaluating $E_{\mathrm{post}}$ at each feasible sampled point, which takes the minimum over all candidate boundary points. This yields a numerical estimate as finite boundary sampling alone does not establish a rigorous safety certificate.



For the damped swing model, the post-reconnection stability region can be characterized by the following energy-based safe set as
\begin{equation}
\mathcal{S}_{\mathrm{post}} := \{(\delta,\omega) \in \mathcal{S}_{\mathrm{adm}}: E_{\mathrm{post}}(\delta,\omega)< E_{\mathrm{crit,post}}\}.
\label{eq:S_post}
\end{equation}
Any reconnection from a state $(\delta,\Delta\omega)\in\mathcal{S}_{\mathrm{post}}$ is thereby certified to remain within the admissible operating boundary under the adopted energy-based boundary criterion.

\subsection{Certified Safe Reconnection Set and Time Windows}
\label{subsec:safe_set}

In addition to the energy-based constraints $\mathcal{S}_{\mathrm{pre}}$ and $\mathcal{S}_{\mathrm{post}}$, reconnection must satisfy operational protection limits. The admissible set defined in \eqref{eq:S_adm} enforces the rotor-angle deviation bounds, the frequency deviation bounds and post-reconnection voltage limits. Thus, the certified safe reconnection region is characterized as the intersection of all three constraint sets as
\begin{equation}
\mathcal{S}_{\mathrm{safe}}
:=
\mathcal{S}_{\mathrm{pre}}\cap\mathcal{S}_{\mathrm{post}}\cap\mathcal{S}_{\mathrm{adm}}.
\label{eq:Ssafe}
\end{equation}

\review{
The certified safe set $\mathcal{S}_{\mathrm{safe}}$ is visualized by discretely sampling the state space $(\delta,\omega)$ over the operating region and retaining those grid points that satisfy the three conditions in \eqref{eq:pre_safe_set}, \eqref{eq:S_post}, and \eqref{eq:S_adm}. The corresponding certified safe reconnection time set is defined as
\begin{equation}
\mathcal T_{\mathrm{safe}}
:=
\left\{
t\ge0:
x^{-}(t)\in
\mathcal S_{\mathrm{safe}}
\right\}.
\label{eq:Tsafe}
\end{equation}
The earliest certified reconnection time is defined as
\begin{equation}
t_{\mathrm{re}}^\star
=
\inf
\mathcal T_{\mathrm{safe}}.
\label{eq:tre_first}
\end{equation}
The infimum is an attained earliest certified reconnection time only if $t_{\mathrm{re}}^\star
\in \mathcal T_{\mathrm{safe}}$.
If $\mathcal T_{\mathrm{safe}}\neq\emptyset$, it can be expressed as the union of one or more disjoint intervals,
\begin{equation*}
\mathcal T_{\mathrm{safe}}
=
\bigcup\nolimits_k I_k,
\end{equation*}
where $I_k$ denotes the $k$th interval, with endpoint inclusion determined by the defining inequalities.}

\review{
Accordingly, reconnection is certified only for
$t_{\mathrm{re}}\in\mathcal T_{\mathrm{safe}}$,
rather than for all
$t_{\mathrm{re}}\ge t_{\mathrm{re}}^\star$.
Because $\dot E_{\mathrm{pre}}\le0$, the set $\mathcal S_{\mathrm{pre}}$ is forward invariant under the waiting dynamics. However, the certified safe reconnection set
$\mathcal S_{\mathrm{safe}}$
is generally not forward invariant under the waiting dynamics,
because its definition also depends on the post-reconnection energy
constraint
$\mathcal S_{\mathrm{post}}$
and the admissibility constraint
$\mathcal S_{\mathrm{adm}}$.
Consequently, the waiting trajectory may leave and subsequently re-enter
$\mathcal S_{\mathrm{safe}}$,which
results in one or more disconnected certified safe reconnection windows.
}


\section{Numerical Study}
\label{sec:results}

The proposed methodology was validated through numerical simulations on a single-machine infinite-bus (SMIB) system with a large-scale DCL, implemented in Julia \cite{Julia-2017} \review{with fault duration $t_f = 0.1\,$s}. The system configuration follows the three-bus model described in Section \ref{sec:model}, with parameters chosen to represent realistic operating conditions for a medium-sized generator supplying a major data center facility as shown in Table \ref{tab:system_parameters}. \Ahmed{The swing equation coefficient $M = 0.0146\,\mathrm{s}^2$ corresponds, for a
60 Hz system, to an inertia constant
$H = M\omega_s/2 \approx 2.75$\,s. This value lies near the lower end of the typical range for synchronous 
generators ($H \approx 1.75$--$10$\,s~\cite{tan2022inertia}), and is therefore representative of a small generating unit or a low-inertia operating condition. Such conditions are consistent with systems experiencing high converter-interfaced generation penetration, where reduced synchronous inertia can make
the reconnection timing decisions more critical. }

The numerical results were generated by discrete sampling of the state of the system. The critical energies $E_{\mathrm{crit,pre}}$ and $E_{\mathrm{crit,post}}$ are computed by boundary optimization and sampling, respectively. A $140 \times 140$ uniform grid over the protection region classifies states based on three criteria: pre-reconnection energy $E_{\mathrm{pre}} < E_{\mathrm{crit,pre}}$, post-reconnection energy $E_{\mathrm{post}} < E_{\mathrm{crit,post}}$, and admissibility constraints, which characterizes the safe reconnection region $\mathcal{S}_{\mathrm{safe}}$. \review{
Trajectory sampling with $\Delta t = 1\ \mathrm{ms}$ resolution along the waiting trajectory identifies the certified safe reconnection time set $\mathcal{T}_{\mathrm{safe}}$ as the collection of time instants for which the system state belongs to $\mathcal{S}_{\mathrm{safe}}$. The earliest certified reconnection time $t_{\mathrm{re}}^\star$ is the first element of $\mathcal{T}_{\mathrm{safe}}$.
}

\begin{table}[t]
\centering
\caption{System Parameters}
\label{tab:system_parameters}
\begin{tabular}{lccc}
\hline
\textbf{Parameter} & \textbf{Symbol} & \textbf{Value} & \textbf{Unit} \\
\hline
\multicolumn{4}{l}{\textit{Generator Parameters}} \\
Inertia coefficient & $M$ & 0.0146 & s$^2$ \\
Damping coefficient & $D$ & 0.05 & s \\
Internal voltage & $e$ & 1.05 & p.u. \\
Transient reactance & $x'_d$ & 0.3 & p.u. \\
Mechanical power & $P_g$ & 0.8 & p.u. \\
\hline
\multicolumn{4}{l}{\textit{Network Parameters}} \\
Pre-fault line reactance & $x_l^{\text{pre}}$ & 0.4 & p.u. \\
Fault-on reactance & $x_l^{\text{f}}$ & 5.0 & p.u. \\
Post-fault line reactance & $x_l^{\text{post}}$ & 0.5 & p.u. \\
\hline
\multicolumn{4}{l}{\textit{DCL Parameters}} \\
Active power & $P_{\text{DCL}}$ & 0.5 & p.u. \\
Reactive power & $Q_{\text{DCL}}$ & 0.2 & p.u. \\
\hline
\multicolumn{4}{l}{\textit{Protection Limits}} \\
Maximum angle deviation & $\delta_{\max}$ & $30$ & degrees \\
Maximum frequency deviation & $\omega_{\max}$ & $2.0$ & rad/s \\
Voltage limits & $V_{\min}, V_{\max}$ & 0.9, 1.1 & p.u. \\
Synchronous frequency & $\omega_s$ & $2\pi \times 60$ & rad/s \\
\hline
\end{tabular}
\end{table}
\subsection{System Equilibrium Points}
\label{subsec:equilibria}

The system equilibria were computed for three operating conditions: pre-fault, no-load (DCL disconnected), and post-fault with DCL reconnected. The results are presented in Table~\ref{tab:equilibrium_points}. \review{
The critical energies are computed as
$E_{\mathrm{crit,pre}}=0.03299$ p.u.
and
$E_{\mathrm{crit,post}}=0.0331$ p.u.
} The pre-reconnection critical energy is dominated by the right boundary, which indicates that escape toward higher rotor angles is the limiting pre-reconnection mechanism. The slight  difference between $E_{\mathrm{crit,pre}}$ and $E_{\mathrm{crit,post}}$ arises from the post-reconnection load connected dynamics, which may be attributed to the voltage-angle coupling.

\begin{table}[t]
\centering
\caption{System Equilibrium Points}
\label{tab:equilibrium_points}
\begin{tabular}{lcccc}
\hline
\textbf{Operating} & \textbf{Rotor Angle} & \textbf{V. Angle} & \textbf{V. Mag.} & \textbf{Electr. Power} \\
\textbf{Condition} & $\delta$ (deg) & $\phi$ (deg) & $V$ (p.u.) & $P_e$ (p.u.) \\
\hline
Pre-fault & 20.66 & 7.08 & 0.9733 & 0.800 \\
No-load & 37.56 & 24.08 & 0.9806 & 0.800 \\
Post-reconnection & 22.54 & 8.90 & 0.9696 & 0.800 \\
\hline
\end{tabular}
\end{table}

\subsection{Certified Safe Reconnection Windows}

\review{
Fig.~\ref{fig:phase_portrait} shows the phase portrait with the certified safe set $\mathcal{S}_{\mathrm{safe}}$ (green region), obtained as the intersection of $\mathcal{S}_{\mathrm{pre}}$ (light blue), $\mathcal{S}_{\mathrm{post}}$ (orange), and $\mathcal{S}_{\mathrm{adm}}$ (light gray). The certified safe set occupies approximately 13.26\% of the admissible region. The waiting trajectory (black line) intersects $\mathcal{S}_{\mathrm{safe}}$ multiple times during its oscillatory evolution.
}

 
\review{
The evolution of the waiting trajectory through the certified safe set gives rise to the certified safe reconnection time set
$\mathcal{T}_{\mathrm{safe}}$.
For the considered operating condition, two disconnected certified reconnection windows are obtained:
\[
0.487 \le t_{\mathrm{re}} \le 0.728~\mathrm{s},
\]
and
\[
1.299 \le t_{\mathrm{re}} \le 1.403~\mathrm{s}.
\]
Accordingly, the earliest detected safe sample is
$t_{\mathrm{re}}^\star=0.487~\mathrm{s}$,
which corresponds to the first point of the certified window. These disconnected windows arise because the waiting trajectory repeatedly enters and leaves the certified safe set during its oscillatory evolution.
}

\begin{figure}[t!]
\centering
\includegraphics[width=0.48\textwidth]{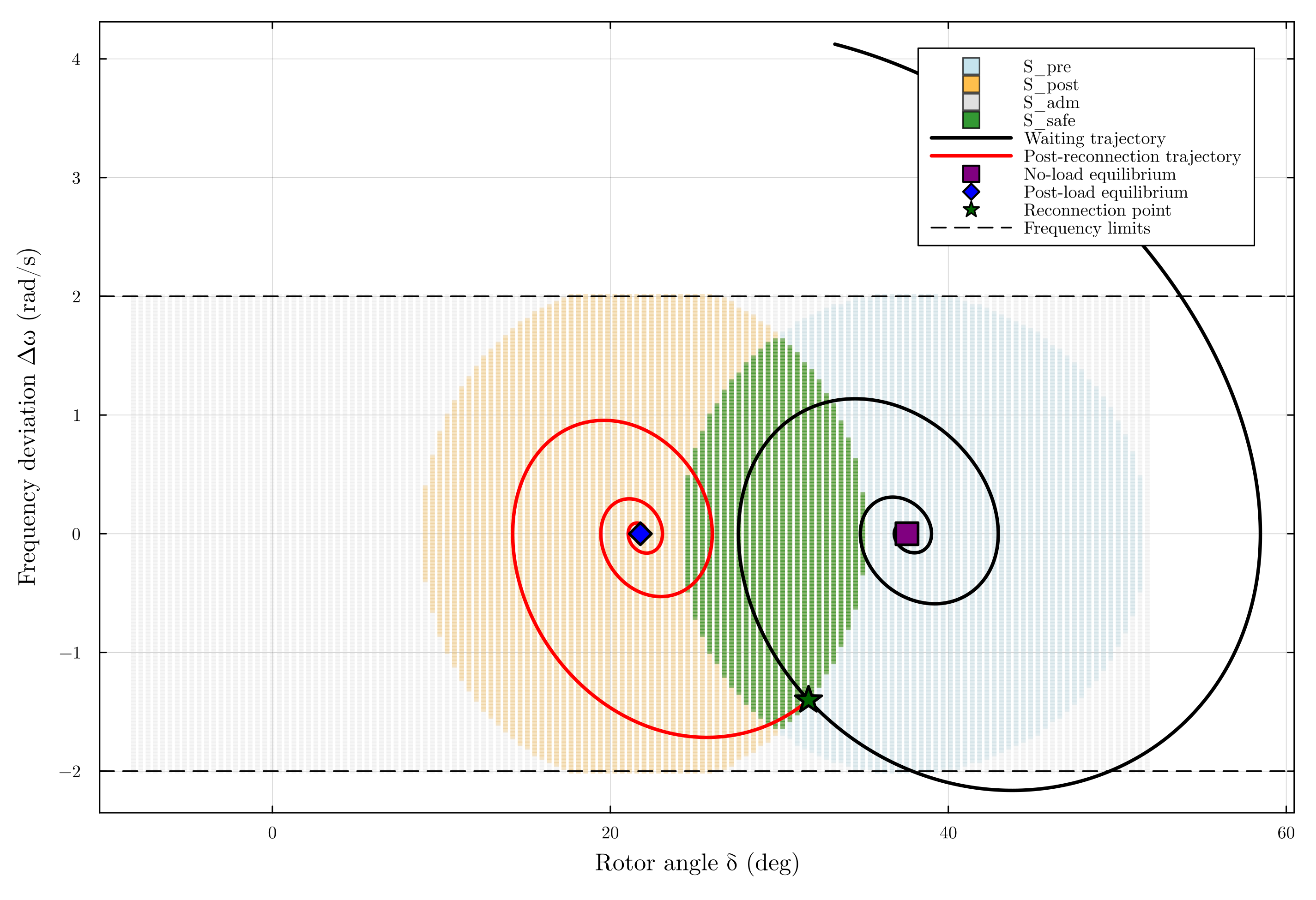}
\caption{Phase portrait showing the certified safe set $\mathcal{S}_{\mathrm{safe}}$, the system trajectories, and the earliest certified reconnection point.}
\label{fig:phase_portrait}
\end{figure}
To validate the importance of proper timing of reconnection, we simulate \review{
reconnection outside the certified safe reconnection windows
}at $t_{\mathrm{re}} = 0.300\ \text{s}$. As shown in Fig.~\ref{fig:unsafe_frequency_deviation}, early reconnection leads to oscillations exceeding protection limits, reaching approximately $3.641$ rad/s, \review{which demonstrates violation of the prescribed operational security constraints.}
\begin{figure}[t!]
\centering
\includegraphics[width=0.48\textwidth]{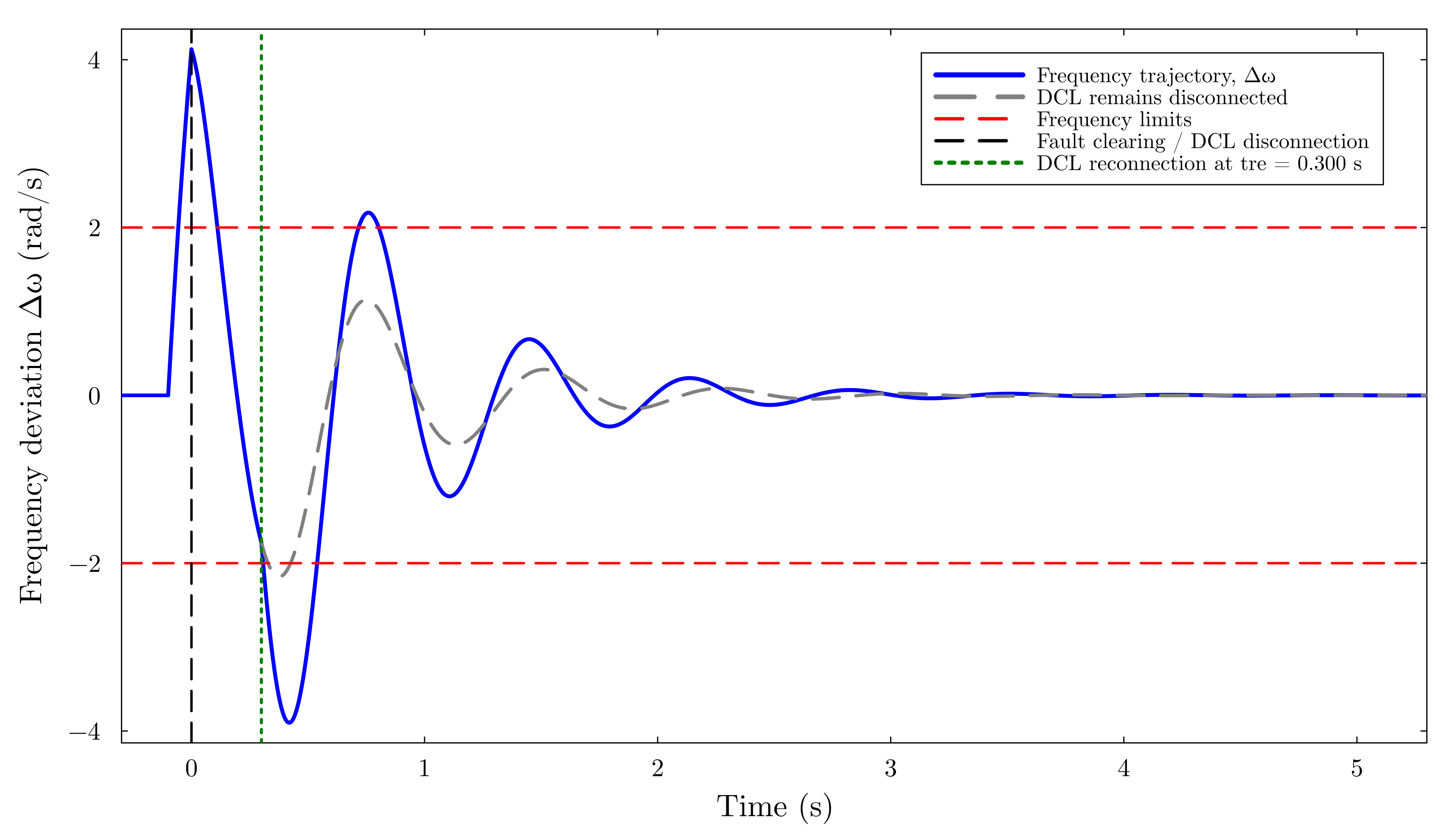}
\caption{Frequency deviation following unsafe reconnection at $t_{\mathrm{re}} = 0.3$ s.}
\label{fig:unsafe_frequency_deviation}
\end{figure}

Fig.~\ref{fig:frequency_deviation} illustrates the frequency deviation during the post-reconnection phase for the safe scenario. After reconnection at $t_{\mathrm{re}}^\star = 0.487\ \text{s}$, the system exhibits damped oscillatory response with initial deviation $\Delta\omega = - 1.41\ \text{rad/s}$, which converge to equilibrium within approximately 3 seconds. The response remains within protection limits ($\pm 2.0\ \text{rad/s}$) throughout the post-reconnection interval.
\begin{figure}[t!]
\centering
\includegraphics[width=0.48\textwidth]{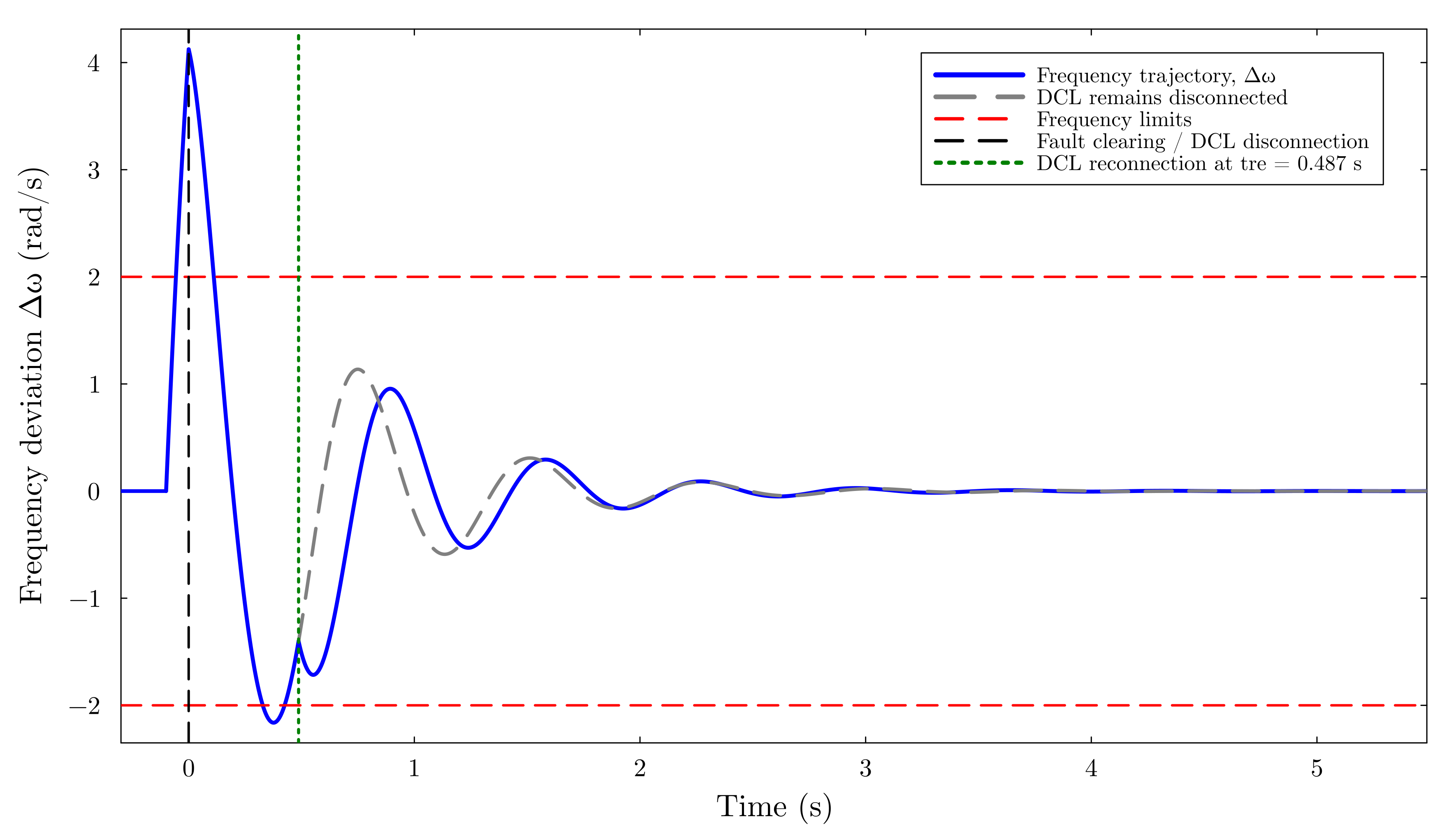}
\caption{Frequency deviation following safe reconnection at $t_{\mathrm{re}} = 0.487$ s.}
\label{fig:frequency_deviation}
\end{figure}

\review{
To further illustrate the certified window characterization, Fig.~\ref{fig:frequency_safe_upper} shows the frequency response for reconnection at
$t_{\mathrm{re}}=0.728$~s, which corresponds to the end point of the first certified reconnection window. The resulting trajectory remains within the prescribed frequency protection limits and converges to the post-reconnection equilibrium. This shows the response is consistent with the first reconnection window.
}
\begin{figure}[t!]
\centering
\includegraphics[width=0.48\textwidth]{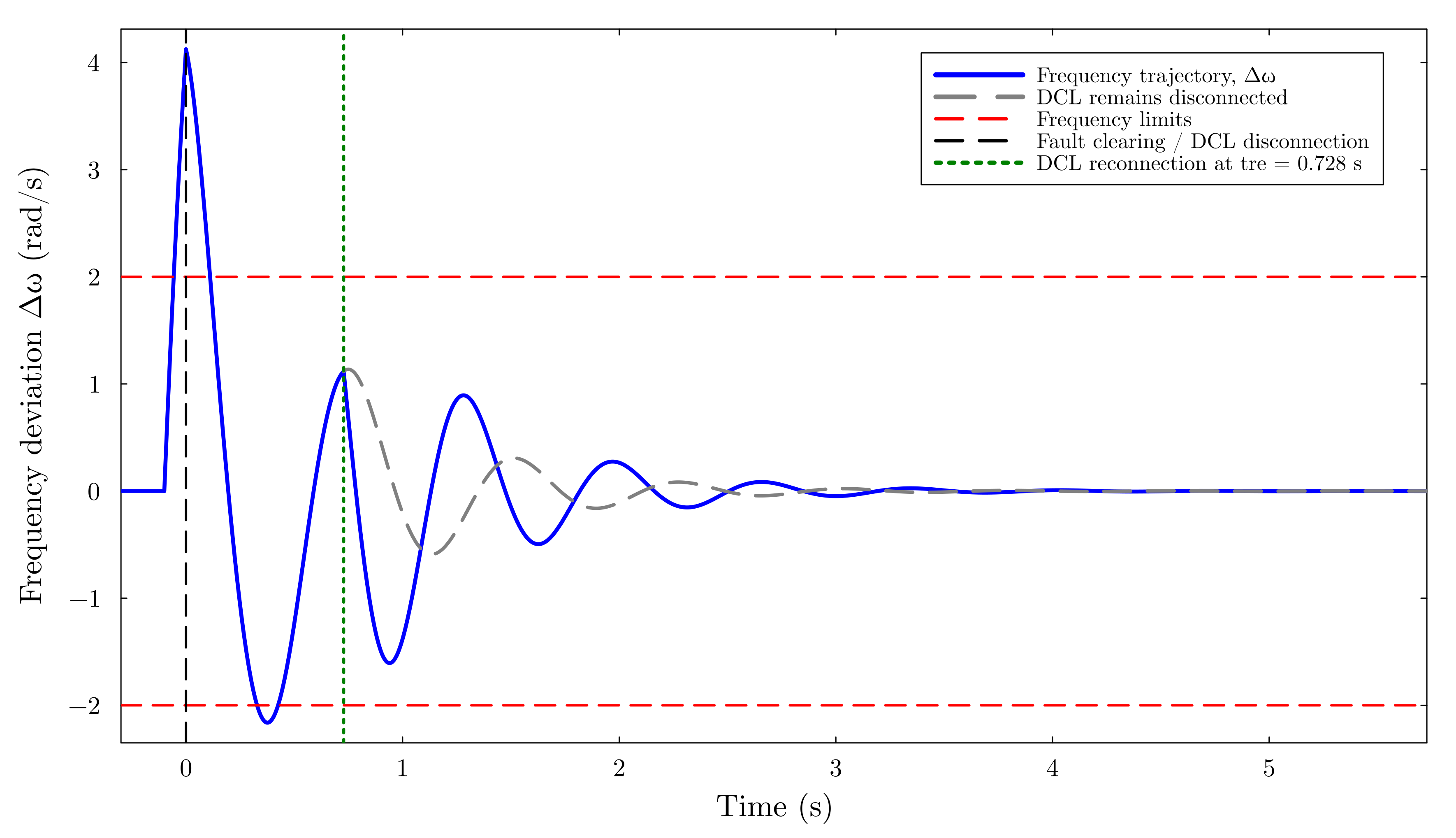}
\caption{Frequency deviation following certified safe reconnection at
$t_{\mathrm{re}}=0.728$~s, corresponding to the right endpoint of the first certified reconnection window.}
\label{fig:frequency_safe_upper}
\end{figure}

\review{
In contrast, reconnection at
$t_{\mathrm{re}}= 0.969$~s occurs outside both certified reconnection windows. As shown in Fig.~\ref{fig:frequency_unsafe_0969}, the resulting oscillation violates the prescribed frequency protection limits which reaches approximately $2.338$ rad/s. Therefore, although
$t_{\mathrm{re}}=0.969$~s is later than the earliest certified time
$t_{\mathrm{re}}^\star=0.487$~s, it is not certified for reconnection. This result demonstrates that the earliest certified reconnection time is not a threshold after which every later reconnection instant is safe.
}
\begin{figure}[t!]
\centering
\includegraphics[width=0.48\textwidth]{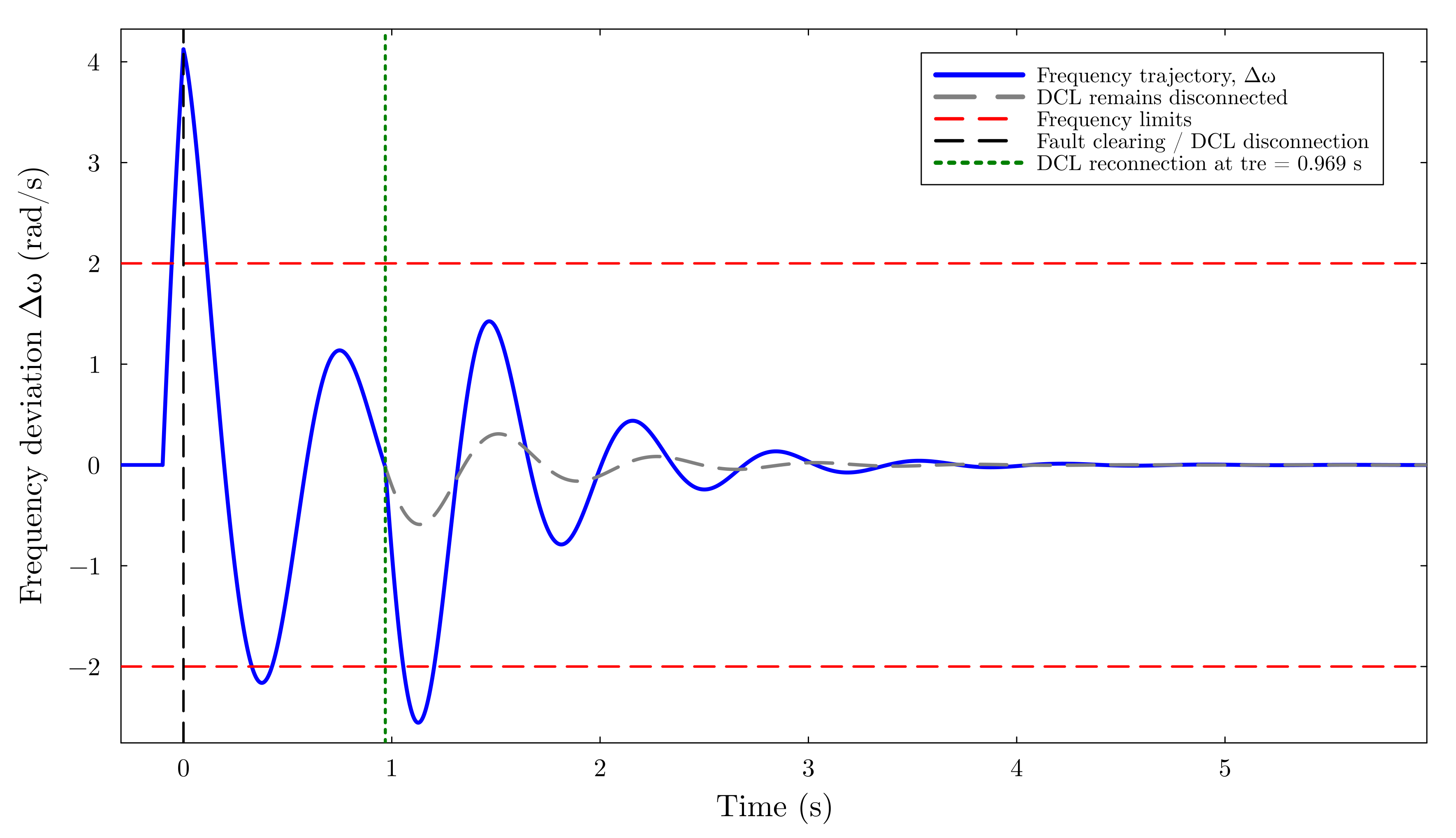}
\caption{Frequency deviation following unsafe reconnection at
$t_{\mathrm{re}}=0.969$~s, which lies outside the certified safe reconnection windows.}
\label{fig:frequency_unsafe_0969}
\end{figure}

\section{Conclusion}
\label{sec:Conclusion}
\review{This paper develops an energy-based analytical framework for determining certified safe reconnection windows for large-scale DCL following grid disturbances. The approach addresses a critical gap in operational decision-making by providing a computationally efficient screening criterion that can inform reconnection policies and coordination between data center UPS controls and system operators. The proposed approach provides sufficient reduced-model conditions ensuring that post-reconnection trajectories satisfy the prescribed operational constraints.}

The key contributions are two-fold: (i) we develop a dual energy function method that combines a pre-reconnection settling condition with post-reconnection safety conditions, which accounts for the voltage-angle coupling introduced by constant power DCL, and  (ii) we formulate a robust boundary optimization for computing critical energy values that define the certified safe reconnection set.

\review{The numerical results demonstrate that the proposed dual-energy-function framework accurately identifies certified safe reconnection windows along the waiting trajectory. The proposed framework successfully distinguishes certified and non-certified reconnection instants without requiring exhaustive post-reconnection time-domain simulations for every candidate reconnection time.}

\Ahmed{Future work will focus on: (i) extending the framework to multi-machine systems and multiple DCLs reconnecting at different times; (ii) validating the method on IEEE benchmark test systems to assess applicability beyond the three-bus model; and (iii) incorporating explicit AVR, Power System Stabilizer (PSS), and governor dynamics to quantify the conservatism of the constant-emf assumption and refine the voltage bounds in the safe set construction.}


\Ahmed{\section*{Acknowledgment}
 A. M. Alkhudaydi would like to thank Royal Commission for Jubail and Yanbu, Saudi Arabia, for the financial support for his postgraduate studies.}

\section{AI Usage Disclosure}
ChatGPT was used to assist with language editing and technical consistency checks. The authors are responsible for the final mathematical arguments, implementation, numerical results, and cited sources.



\balance
\bibliographystyle{ieeetr}
\bibliography{ref_updated}

@techreport{cru_21_124,
  author      = {{Commission for Regulation of Utilities}},
  title       = {Direction to the System Operators Related to Data Centre Grid Connection Processing},
  institution = {Commission for Regulation of Utilities},
  type        = {Decision Paper},
  number      = {CRU/21/124},
  address     = {Dublin, Ireland},
  month       = nov,
  year        = {2021},
  url         = {https://cruie-live-96ca64acab2247eca8a850a7e54b-5b34f62.divio-media.com/documents/CRU21124-CRU-Direction-to-the-System-Operators-related-to-Data-Centre-grid-connection-.pdf}
}

@book{kundur_pssc,
  author    = {Kundur, Prabha S.},
  title     = {Power System Stability and Control},
  publisher = {McGraw-Hill},
  address   = {New York, NY, USA},
  year      = {1994},
  isbn      = {9780070359581}
}

@book{pai_energy_function,
  author    = {Pai, M. A.},
  title     = {Energy Function Analysis for Power System Stability},
  publisher = {Springer},
  address   = {New York, NY, USA},
  year      = {1989},
  isbn      = {9780792390350},
  doi       = {10.1007/978-1-4613-1635-0}
}

@book{sauer_pai_chow_2017,
  author    = {Sauer, Peter W. and Pai, M. A. and Chow, Joe H.},
  title     = {Power System Dynamics and Stability: With Synchrophasor Measurement and Power System Toolbox},
  publisher = {Wiley-IEEE Press},
  edition   = {2},
  year      = {2017},
  isbn      = {9781119355779},
  doi       = {10.1002/9781119355755}
}

@techreport{IEA_Electricity2024,
  author      = {{International Energy Agency}},
  title       = {{Electricity 2024}: Analysis and Forecast to 2026},
  institution = {International Energy Agency},
  type        = {Report},
  month       = jan,
  year        = {2024},
  url         = {https://www.iea.org/reports/electricity-2024}
}

@techreport{IEA_EnergyAI,
  author      = {{International Energy Agency}},
  title       = {Energy and {AI}},
  institution = {International Energy Agency},
  type        = {Report},
  month       = apr,
  year        = {2025},
  url         = {https://www.iea.org/reports/energy-and-ai}
}

@article{Guerrero2007,
  author  = {Guerrero, Josep M. and de Vicu{\~n}a, Luis Garcia and Uceda, Javier},
  title   = {Uninterruptible power supply systems provide protection},
  journal = {IEEE Ind. Electron. Mag.},
  volume  = {1},
  number  = {1},
  pages   = {28--38},
  month   = mar,
  year    = {2007},
  doi     = {10.1109/MIE.2007.357184}
}

@article{Guerrero2008,
  author  = {Guerrero, Josep M. and Hang, Lijun and Uceda, Javier},
  title   = {Control of distributed uninterruptible power supply systems},
  journal = {IEEE Trans. Ind. Electron.},
  volume  = {55},
  number  = {8},
  pages   = {2845--2859},
  month   = aug,
  year    = {2008},
  doi     = {10.1109/TIE.2008.924173}
}

@standard{IEC62040_3_2021,
  organization = {{International Electrotechnical Commission}},
  title        = {{IEC 62040-3:2021}: Uninterruptible power systems ({UPS}) -- Part 3: Method of specifying the performance and test requirements},
  number       = {IEC 62040-3:2021},
  month        = apr,
  year         = {2021},
  url          = {https://webstore.iec.ch/en/publication/60140}
}

@standard{IEEE1159_2019,
  organization = {{IEEE}},
  title        = {{IEEE Std 1159-2019}: Recommended Practice for Monitoring Electric Power Quality},
  number       = {IEEE Std 1159-2019},
  month        = aug,
  year         = {2019},
  url          = {https://standards.ieee.org/standard/1159-2019.html}
}

@inproceedings{kerci_cigre_c4_10907_2024,
  author    = {K{\"e}r{\c{c}}i, T. and Duggan, C. and Farooq, U. and Tweed, S. and Val Escudero, M.},
  title     = {Impact of converter-based demand on frequency quality in the {Ireland} and {Northern Ireland} power systems},
  booktitle = {Proc. {CIGRE} Paris Session},
  number    = {C4-10907-2024},
  year      = {2024},
  url       = {https://www.e-cigre.org/publications/detail/c4-10907-2024-impact-of-converter-based-demand-on-frequency-quality-in-the-ireland-and-northern-ireland-power-systems.html}
}

@misc{jimenez_ruiz_milano_dc_transient_2025,
  author        = {Jim{\'e}nez-Ruiz, Alberto and Milano, Federico},
  title         = {Data Center Model for Transient Stability Analysis of Power Systems},
  howpublished  = {arXiv:2505.16575},
  archivePrefix = {arXiv},
  eprint        = {2505.16575},
  primaryClass  = {eess.SY},
  month         = may,
  year          = {2025},
  doi           = {10.48550/arXiv.2505.16575},
  url           = {https://arxiv.org/abs/2505.16575}
}

@article{ieee_tf_standard_load_models_1995,
  author  = {Price, W. W. and Taylor, C. W. and Rogers, G. J.},
  title   = {Standard load models for power flow and dynamic performance simulation},
  journal = {IEEE Trans. Power Syst.},
  volume  = {10},
  number  = {3},
  pages   = {1302--1313},
  month   = aug,
  year    = {1995},
  doi     = {10.1109/59.466523}
}

@article{ElengaBaningobera2025FFR_WAMS,
  author  = {Elenga Baningobera, Bwandakassy and Oleinikova, Irina and Uhlen, Kjetil},
  title   = {Optimizing frequency stability with adaptive fast frequency reserves and {Wide-Area Monitoring Systems}},
  journal = {Int. J. Electr. Power Energy Syst.},
  volume  = {171},
  number  = {},
  month   = oct,
  year    = {2025},
  note    = {{Art. no. 110951}},
  doi     = {10.1016/j.ijepes.2025.110951}
}

@article{Saha2023RES_FrequencyBehaviour,
  author  = {Saha, S. and Saleem, M. I. and Roy, T. K.},
  title   = {Impact of high penetration of renewable energy sources on grid frequency behaviour},
  journal = {Int. J. Electr. Power Energy Syst.},
  volume  = {145},
  number  = {},
  pages   = {1--13},
  month   = feb,
  year    = {2023},
  note    = {{Art. no. 108701}},
  doi     = {10.1016/j.ijepes.2022.108701}
}

@article{tan2022inertia,
  author  = {Tan, Bendong and Zhao, Junbo and Netto, Marcos and Krishnan, Venkat and Terzija, Vladimir and Zhang, Yingchen},
  title   = {Power system inertia estimation: Review of methods and the impacts of converter-interfaced generations},
  journal = {Int. J. Electr. Power Energy Syst.},
  volume  = {134},
  number  = {},
  year    = {2022},
  note    = {{Art. no. 107362}},
  doi     = {10.1016/j.ijepes.2021.107362}
}

@article{Julia-2017,
    title={Julia: A fresh approach to numerical computing},
    author={Bezanson, Jeff and Edelman, Alan and Karpinski, Stefan and Shah, Viral B},
    journal={SIAM {R}eview},
    volume={59},
    number={1},
    pages={65--98},
    year={2017},
    publisher={SIAM},
    doi={10.1137/141000671},
    url={https://epubs.siam.org/doi/10.1137/141000671}
}
\balance

\endgroup
\end{document}